\documentclass[runningheads]{svmult}
\usepackage{amssymb}   
\usepackage{makeidx}   
\usepackage{graphicx}  
\usepackage{subeqnar}  
\usepackage{multicol}  
\usepackage{cropmark} 
\usepackage{physprbb}  
\begin{document}
\title*{Leptogenesis in a Prompt Decay Scenario}
\toctitle{Leptogenesis in a Prompt Decay Scenario}
%
%
\titlerunning{Leptogenesis in a Prompt Decay Scenario}
%
\author{Lu{\'{\i}}s Bento}
\authorrunning{Lu{\'{\i}}s Bento}
%
%
\institute{Centro de F\'{\i}sica Nuclear da Universidade de Lisboa\\
Avenida Prof. Gama Pinto 2, 1649-003 Lisboa, Portugal }

\maketitle              

\begin{abstract}
Leptogenesis is studied within the seesaw neutrino mass model in a regime where all sterile neutrinos have prompt rather than delayed decays.
It is shown that during neutrino thermal production lepton asymmetries are generated in both active lepton and sterile neutrino sectors.
The large $B-L$ asymmetry is slowly pumped into the chemically decoupled right-handed quarks and leptons and baryon number sector which later protect $B-L$ from fast $L$ violating processes.
The dependence of the final baryon asymmetry on couplings and masses is totally different from the decay scenario.
$B$ does not vanish in the limit of degenerate light neutrinos and the observed asymmetry is naturally obtained for a sum of square masses, $\bar{m}^2$, between the atmospheric neutrino mass gap and $0.2 \,\mathrm{eV}^2$.
\end{abstract}

\section{Introduction}

Leptogenesis is an attractive way of generating the baryon number of the 
Universe~\cite{Buchmuller00}. 
The main idea put forward by Fukugita and Yanagida~\cite{Fukugita86}
is that a lepton number asymmetry can be produced in the decays of heavy sterile neutrinos into leptons and Higgs bosons and such an asymmetry 
is partially transferred to the baryon sector through electroweak sphaleron processes~\cite{Kuzmin85}
that violate $B$ and $L$ but not $B-L$.
The mechanism requires lepton number and $CP$ violation, ensured by 
neutrino Majorana masses and
complex Yukawa couplings.
Both masses and couplings form the well known 
seesaw model~\cite{Gellmann79}
of light neutrino masses and thus establish a close relation between baryogenesis and 
low energy phenomenology.

The third Sakharov condition~\cite{Sakharov67}, 
the departure from thermal equilibrium, is satisfied in the delayed decay scenario.
It means here that the Majorana neutrinos live longer than the age of the Universe when they approach non-relativistic temperatures and their production is Boltzmann suppressed.
It turns out that such condition puts an upper bound on the natural order of magnitude of the light neutrino masses~\cite{Fischler91,Buchmuller00}
well below the mass scale indicated by the atmospheric neutrino 
anomaly~\cite{skatm00}.
This natural bound can be circumvented to some extent by fine-tuning the parameters  to maximize the produced $CP$ asymmetry but even so a quasi-degenerate light neutrino spectrum is excluded~\cite{davidson02,Buchmuller02}.

As the seesaw provides a simple and elegant neutrino mass mechanism, it would be nice to have a naturally successful leptogenesis mechanism in the case where all heavy neutrinos have prompt rather than delayed decays.
Given that the heavy neutrino Majorana masses and Yukawa couplings provide the necessary $B-L$ and $CP$ violations the main concern is to satisfy the out-of-equilibrium condition.
This is naturally achieved by assuming that the heavy sterile neutrinos do not couple to the inflaton field and are only thermally produced during the radiation era from standard lepton and Higgs particles.
Before the heavy neutrinos reach thermal equilibrium abundances a non-zero $B-L$ asymmetry can in principle be generated.
That is the subject of this paper. 

In the next section we describe the mechanism in more detail and in the following ones we present the main results concerning first, the generation of the primary $B-L$ asymmetry and second, the protection mechanism that prevents $B-L$ from being completely washed out when the lepton number violating reactions reach thermal equilibrium.
In the last section we summarize the mechanism and make a brief discussion.

\section{Thermal production - prompt decay scenario}

The seesaw mechanism adds to the standard model singlet (left-handed) neutrinos,
$N_a$,
with heavy Majorana masses and Yukawa couplings to the standard lepton and Higgs doublets,
$l_i$, $\phi$, of the form
\begin{equation}\label{Yuk} 
h_{ia}l_i N_a \phi + \frac{1}{2} M_{a} N_a N_a  
+  {\rm H.C.}  \; .
\end{equation}
Spontaneous breaking of SU(2) $\times$ U(1) yields the light neutrino mass matrix
($v= \langle \phi^0 \rangle$)
\begin{equation}\label{mij}
m_{ij} =- (h M^{-1} h^{T} )_{ij} \, v^{2}
			\; .
\end{equation}
The proper decay rate of $N_a$ into leptons and Higgs is
$\Gamma_a = (h ^{\dagger} h)_{aa} M_a /8\pi$.
Delayed decay occurs when the ratio to the Hubble expansion rate, 
$K_a = \Gamma_a /H$,
is small at the temperature $T=M_a$.
In the radiation era, 
$H = 1.66\, g_\ast^{1/2}T^2/M_{P}$, 
where $g_\ast$ denotes the number of relativistic degrees 
of freedom, 107.5 in the standard model.
It is enough to compare the sum
\begin{equation}\label{k}
K = \sum K_a = 
\frac{\mathrm{tr}[h M^{-1} h ^{\dagger}] v^2}{10^{-3} \; \mathrm{eV}}
\end{equation}
with the light neutrino scale 
$\mathrm{tr} [ m ]$
to conclude that the delayed decay condition, $K_a <1$, is in general in conflict with the atmospheric neutrino mass gap~\cite{skatm00},
$\Delta m^2 \approx 3 \cdot 10^{-3} \,\mathrm{eV}^2$,
which implies $K > 50$. 
Strictly speaking, the delayed decay scenario only requires that the lightest of the heavy neutrinos satisfies $K_a <1$, and larger $K_a$ values are tolerated for a certain choice of parameters that maximize the $CP$ 
asymmetry~\cite{davidson02,Buchmuller02}. 
But that is not the most natural picture,
in particular if the light neutrinos are 
quasi-degenerate~\cite{davidson02,Buchmuller02}.

As it turns out, there is a simple modification of the original leptogenesis mechanism capable of realizing the out-of-equilibrium condition and producing a large lepton asymmetry.
According to the inflation paradigm~\cite{Kolb90}, 
the radiation era starts when a scalar field, so-called inflaton, decays and transforms its energy into the form of particles so constituting the initial thermal bath. 
If sterile neutrinos do not couple to the inflaton,
 they can be only thermally produced from standard lepton and Higgs particles through Yukawa interactions.
These processes are naturally out-of-equilibrium at very high  temperatures and it takes some time before sterile neutrinos reach thermal equilibrium densities.
During that period there is a clear departure from equilibrium and a $B-L$ asymmetry can be generated.

In this scenario a primary lepton asymmetry is generated in the lepton doublet and sterile neutrino sectors.
A fraction of it is transferred to the bottom $b_R$, tau $\tau_R$, charm $c_R$ and other right-handed fermions through slow Yukawa interactions.
Another fraction is converted into baryon number by weak sphaleron processes.
It is crucial for the survival of such baryon number and 
$b_R$, $\tau_R$, $c_R$, ...
asymmetries that the heavy Majorana neutrinos decay and vanish from the Universe before the weak sphalerons or $b_R$, $\tau_R$ ($c_R$) Yukawa couplings enter in equilibrium, which happens at a temperature around~\cite{Bento03} 
$10^{12} \,\mathrm{GeV}$ 
($10^{11} \,\mathrm{GeV}$).
By that time the $\Delta L=2$ reactions mediated by off-shell Majorana neutrinos should also be out-of-equilibrium.
In other words, the moment when the $\Delta L=2$ reactions freeze out determines which decoupled sectors or quantum numbers protect and store a fraction of the primary $B-L$ asymmetry.
Their rate relates to the sum of light neutrino square masses
$\bar{m}^2 =\mathrm{tr} [ m m^\dagger ] $ as
$\Gamma_2 \approx  \bar{m}^2 T^3 /(\pi^3 v^{4})$~\cite{Fukugita90,Harvey90}.
They are out of equilibrium, i.e., 
$\Gamma_2 /H < 1$, at temperatures below
\begin{equation}
T_\mathrm{out} \approx \frac{0.04 \,\mathrm{eV}^2}{\bar{m}^2} \, 
10^{12} \,\mathrm{GeV}
	\; . 					\label{tout}
\end{equation}
We assume of course that the masses $M_a$ are larger than 
$T_\mathrm{out}$ so that the heavy neutrinos vanish above that temperature.
$T_\mathrm{out} \approx 10^{12} \,\mathrm{GeV}$ ($10^{11} \,\mathrm{GeV}$)
for neutrino masses $\bar{m}^2 = 0.04$ ($0.4 \,\mathrm{eV}^2$),
well above the atmospheric neutrino mass gap.
As will be shown below, if $\bar{m}^2 \lesssim 0.2 \,\mathrm{eV}^2$ then the baryon number sector and the right-handed quark $b_R$ have the dominant role in protecting $B-L$ and producing the present baryon number of the Universe.
If $\bar{m}^2$ exceeds $0.2 \,\mathrm{eV}^2$ that role would be played by the charm quark $c_R$ but not efficiently enough to produce the present baryon asymmetry.

\section{Generation of $B-L$}

We assume that the Universe is initially empty of sterile neutrinos,
not produced in the inflaton decay
but only thermally from standard leptons and Higgs.
For definiteness we assume a reheating temperature much higher than the neutrino masses.
The dominant production reaction is the Higgs boson decay into leptons and neutrinos, $\bar{\phi} \rightarrow l_i N_a$,
allowed by a Higgs thermal mass, 
$m_\phi = x_\phi T \sim 0.6 \, T$~\cite{Cline93},
much higher than the lepton thermal masses.
The Yukawa top quark scattering processes like
$q_t \bar{t}_R \rightarrow l_i N_a$
and other channels contribute to not more than 40 \% of the total $N_a$ production rate.
The Higgs boson proper decay rate into $N_a$ is 
$\Gamma_\phi^a = (h ^{\dagger} h)_{aa} m_\phi /16\pi$
per isospin state.
In the radiation era the initially zero neutrino densities $n_a$ converge to the fermion equilibrium density
$n_\mathrm{eq} = 0.90\, T^3/\pi^2$  
with relaxation temperatures
$T_a \approx 0.35\, K_a M_a$:
\begin{equation}\label{na}
n_a = n_\mathrm{eq} \left( 1 - e^{-T_a /T} \right)		\; .
\end{equation}
As a consequence of the prompt decay assumption, $K_a \gtrsim 70$, sterile neutrinos reach thermal equilibrium while they are still ultra relativistic.

\begin{figure}[b]
\begin{center}
\includegraphics[width=.65\textwidth]{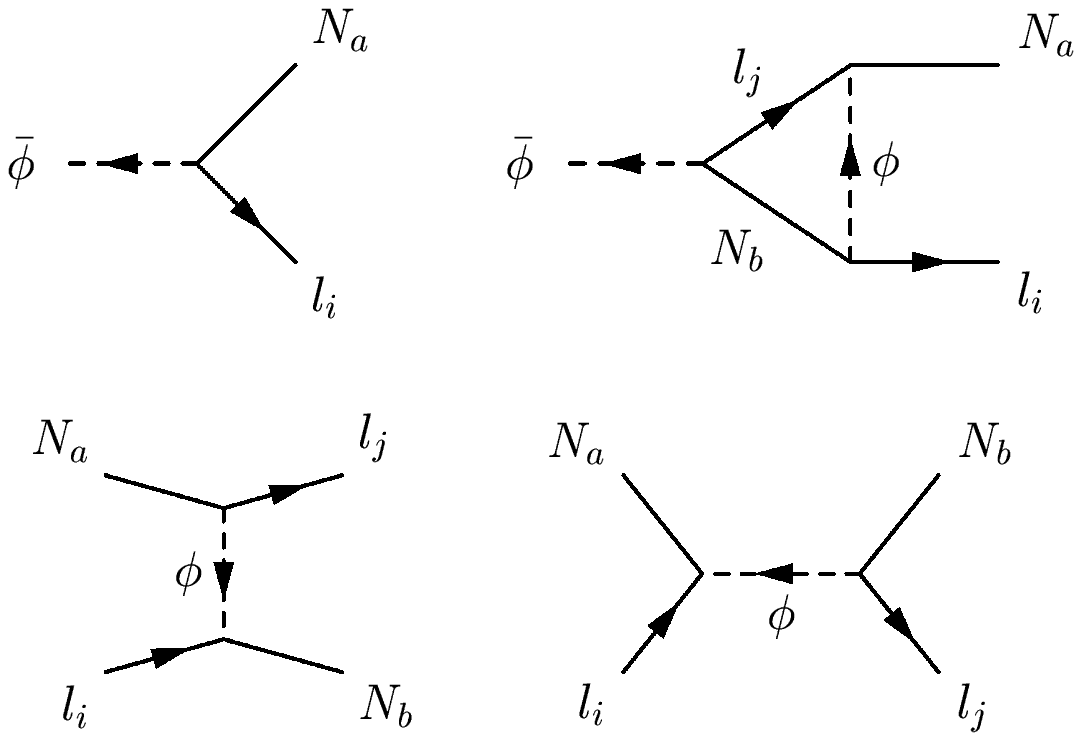}
\end{center}
\caption[]{Diagrams contributing to the CP-asymmetries in decays, inverse decays and scatterings}
\label{fig1}
\end{figure}

Leptogenesis is dominated by the following $CP$ asymmetric reactions:
Higgs decays into leptons and sterile neutrinos, inverse decays and scatterings of leptons off neutrinos.
The asymmetries result from the interference between the diagrams on the left side of Figure \ref{fig1} with the absorptive part of the respective diagram on the right-hand side.
Unitarity and $CPT$ invariance impose constraints on the reaction rates such that when all particles are in thermal equilibrium the asymmetry source terms cancel each other~\cite{Weinberg79,Kolb80,bento02}.
Generically denoting
the difference between the total rates of a reaction 
$X \rightarrow Y$ and its $CP$-conjugate $\bar{X} \rightarrow \bar{Y}$
as $\Delta \gamma (X \rightarrow Y)$
(in a fixed comoving volume),
the relevant constraints are expressed in terms of thermal equilibrium rates 
as:
\begin{subeqnarray}
\Delta \gamma (\bar{\phi} \rightarrow N_a l_i) _\mathrm{eq} + \Delta \gamma  ( N_a l_i \rightarrow \bar{\phi} ) _\mathrm{eq} =0
 	\;, \label{CPTa} 	\\
\Delta \gamma ( N_a l_i \rightarrow  N_b l_j )_\mathrm{eq} + \Delta \gamma ( N_b l_j  \rightarrow N_a l_i ) _\mathrm{eq} =0
 	\;, \label{CPTb}	\\
\Delta \gamma  ( N_a l_i \rightarrow \bar{\phi} )_\mathrm{eq} + \sum_{bj} \Delta \gamma  ( N_a l_i \rightarrow  N_b l_j  ) _\mathrm{eq} =0
 	\;  \label{CPTc} .	
\end{subeqnarray}

The evolution of the set of observables $Q_\alpha$ is governed by the  Boltzmann equations, whose general structure is,
integrating over a constant comoving volume,
$\dot{Q}_\alpha = - \Gamma_{\alpha \beta} Q_\beta + (\dot{Q}_\alpha)_S $.
The first are transport terms and $(\dot{Q}_\alpha)_S$ stand for source terms.
The particle asymmetry sources vanish in thermal equilibrium 
but, as long as the $N_a$ neutrinos stay rarefied, scatterings and inverse decays do not match Higgs decays and particle asymmetries develop in the various lepton flavors, sterile neutrinos, and Higgs boson as well, as enforced by hypercharge conservation.
The calculation of the source terms is simplified by employing Boltzmann rather than quantum statistics as is common practice.
Relating the out-of-equilibrium and thermal equilibrium reaction rates as $\gamma = \gamma_\mathrm{eq} \, n_a/n_\mathrm{eq} $
for $N_a X \rightarrow Y$, and  $\gamma = \gamma_\mathrm{eq} $
for $X \rightarrow N_a Y$,
one derives from the above $CPT$ and unitarity constraints the $l_i$ lepton number sources,
\begin{eqnarray}\label{lis}
(\dot{L}_i )_S
= \sum_{abj} \frac{n_a - n_\mathrm{eq}}{n_\mathrm{eq}}
\, \Delta \gamma (N_a l_j\rightarrow  N_b l_i )_\mathrm{eq} 
	\;,	\\
 \Delta \gamma (N_a l_j\rightarrow  N_b l_i )_\mathrm{eq} = 
 - \frac{c_l}{32 \pi^2} \frac{Y_\phi}{T} J_{ijab }
	\;,	
\end{eqnarray}
where $c_l \approx 0.1$,
$J_{ijab} = \mathrm{Im} \{ h_{ia} h_{ja} h_{ib}^\ast h_{jb}^\ast  \} M_a M_b$
and $Y_\phi$ is the Higgs abundance per isospin state.

Sterile neutrinos also have lepton number, $L=-1$ for left-handed $N_a$ fields and $L=+1$ for the conjugate $\bar{N}_a$.
But what matters for particle propagation and transport of lepton number are the spin eigenstates rather than chiral states.
When neutrinos are ultra relativistic, helicity and chiral states are almost identical, but
when they are non-relativistic Majorana masses take over and lepton number vanishes.
In fact, a positive (negative) helicity eigenstate has a well defined lepton number expectation value equal to the neutrino speed $u$ ($-u$).
Helicity is not an invariant quantity, however, there is a privileged reference frame, the comoving thermal bath rest frame.
In that frame, isotropy ensures that the spin density matrix is diagonal in the helicity basis.
This is not true in any other frame, and means that neutrinos can be divided in two populations of opposite helicities.

Under the non-equilibrium conditions of the leptogenesis era the two neutrino helicity populations develop unequal abundances and net lepton numbers in much the same way as the standard leptons.
From the same decay, inverse decay and scattering processes one derives the $N_b$ lepton number source terms,
\begin{equation}\label{lbs}
(\dot{L}_b)_S = - r_N \sum_{aij} \frac{n_a - n_\mathrm{eq}}{n_\mathrm{eq}}
\, \Delta \gamma (N_a l_j\rightarrow  N_b l_i )_\mathrm{eq} 
	\;,	
\end{equation}
where $r_N$ is a coefficient estimated to be of the order but larger than unit.
Summing both lepton and neutrino contributions, one obtains the
total $B-L$ asymmetry as a function of the temperature:
\begin{equation}\label{BmL}
B-L = c\, Y_f \sum_{ab} 
\frac{\mathrm{Im} [(h^\dagger h)_{ba}]^2}{(h^\dagger h)_{aa}}
\frac{M_a M_b}{ T_a^{2}}
f\left(\frac{T_a}{T} \right) ,
\end{equation}
where $c \approx 2 c_l (r_N -1)/\pi \sim 0.1$, 
$Y_f $
is the massless fermion thermal abundance
and $f(x)= 2-(x^2 +2x +2 ) e^{-x}$.

The other question is how fast Majorana masses erase lepton number.
In a decay $\bar{\phi} \rightarrow l_i N_a$, lepton number suffers a variation 
$\Delta L = 2$ if $N_a$ has positive helicity,
which occurs with a branching fraction of about $10\, M_a^2/T^2 $.
It makes an average variation $\Delta L \approx 20\,M_a^2/T^2$ per decay
and the $N_a$ production rates through Higgs decays are equal to
$H \, T'_a /T$ with  
$T'_a \approx \frac{1}{6} K_a M_a$.
Thus, lepton number violation enters in equilibrium at a temperature 
$T_\mathrm{in}  \approx \sum 20\, T'_a M_a^2 /T_\mathrm{in} ^2$,
i.e.,
\begin{equation} 
T_\mathrm{in} ^3 \approx  \frac{10}{3} \sum K_a M_a^3
	\; .									\label{tin}
\end{equation}
The $\Delta L = 2$ scattering processes mediated by Majorana neutrinos like
$l_i \phi \rightarrow \bar{l}_j \bar{\phi }$,
that are so important when the Majorana neutrinos are non relativistic and even at lower temperatures when they are off-shell, are completely negligible in comparison with the Higgs boson decays $\Delta L =2$ transitions at relativistic temperatures.
The relative strength is of the order of $h_{ia}^2 /200$.
For constants $K_a$ as large as 200, the sterile neutrinos enter in thermal equilibrium at temperatures 
$T_a \approx 0.35\, K_a M_a \sim 70 M_a$ 
but the Majorana masses, i.e., the $\Delta L \neq 0$ transitions only enter in equilibrium at 
$T_\mathrm{in}  \sim 9\, M_3 \sim T_a /8$,
where $M_3$ denotes the largest of the Majorana masses.
$T_\mathrm{in} $ is quite smaller than the relaxation temperatures $T_a$ and the larger the $K_a$
constants are the smaller is $T_\mathrm{in} $ in comparison with $T_a$.

\section{Protecting $B-L$}

At temperatures above $T_\mathrm{in} $, $B-L$ is not significantly damped and is slowly transferred via Yukawa interactions to the right-handed leptons and quarks, $e_i$, $u_i$ and $d_i$
(with the exception of the top, $t_R$, very early in chemical equilibrium).
The dominant processes are Higgs boson decays into 
leptons~\cite{Cline93}, 
$\bar{\phi} \leftrightarrow e_i \bar{l}_i$,
kinematically forbidden in the case of quarks due to comparable quark thermal masses~\cite{Davidson94}, 
and scatterings like
$\bar{\phi} t_R \leftrightarrow b_R \phi$,
$ b_L t_L \leftrightarrow b_R t_R$.
These processes generate the otherwise zero asymmetries 
$Q_{e_i} \equiv L_{e_i}$
and $Q_{u_i} = B_{u_i} - B_{u_R}$, $Q_{d_i} = B_{d_i} - B_{u_R}$.
For right-handed quarks,
the contrast asymmetries with respect to the up quark $u_R$ are the appropriate charges because, contrary to the individual flavor asymmetries, they are conserved by QCD instantons~\cite{McLerran91} and are only violated by Yukawa couplings.
Their growth rates are determined by the total Yukawa production rates, $\Gamma_\chi Y_f$, of
$\chi =e_i$, $u_i = c_R$, $d_i$, and degeneracy parameters $\eta = \mu/T$ as,
\begin{subeqnarray}
\dot{Q}_{e_i} = 2 (\eta_{l_i} - \eta_{\phi} - \eta_{e_i})\,
\Gamma_{e_i} Y_f
		\; ,		\label{qei}				\\
\dot{Q}_{u_i} = 2 (\eta_{q_i} + \eta_{\phi} - \eta_{u_i})\,
\Gamma_{u_i} Y_f
		\; ,			\label{qui}	\\
\dot{Q}_{d_i} = 2 (\eta_{q_i} - \eta_{\phi} - \eta_{d_i})\,
\Gamma_{d_i} Y_f
		\;,			\label{qdi}
\end{subeqnarray}
neglecting flavor violation and  $u_R$ Yukawa coupling.
$\Gamma_\chi$ scale as the temperature and the ratios $\Gamma_\chi /H = T_\chi /T$ 
define equilibrium temperatures, $T_\chi$, under which the respective detailed balance equations are enforced.
We estimate $T_b \approx 2 \cdot 10^{12} \,\mathrm{GeV}$
and $T_{\tau} \approx 10^{12} \,\mathrm{GeV}$.
On the other hand, weak sphalerons transfer part of the $B-L$ asymmetry to the baryon sector and the baryon number growth 
rate~\cite{Khlebnikov88,Bochkarev87,Dine90,Bento03}
in a volume $V$ is given by
\begin{eqnarray}	\label{dBdt}
	\frac{dB}{dt} = - \frac{3}{2}\Gamma_\mathrm{sph}  \,V \sum_i (3 \eta_{q_i} + \eta_{l_i})
		\;,																				
\end{eqnarray}
where $\Gamma_\mathrm{sph} $ is the sphaleron diffusion rate.

The integration of the above equations 
requires knowing the relations between the chemical potentials and the leptogenesis sources given in equations (\ref{lis}) and (\ref{lbs}). 
In general there is no direct relation because the partial particle asymmetries are not conserved due to flavor changing Yukawa couplings.
As soon as the first neutrino, say $N_c$, is in equilibrium ($T < T_c$), the heavy neutrino and standard lepton flavors are rapidly violated 
enforcing their chemical potentials $\eta_{l_i}$ to be identical: 
$\eta_{l_i}+\eta_{N_c}+\eta_{\phi}=0$.
The Higgs communicates also its asymmetry to $t_R$ and top quark doublet:
$\eta_{q_t}+\eta_{\phi}-\eta_{t_R}=0$.
Because the $Q$ asymmetries increase rapidly with time, initially as $T^{-4}$,
we concentrate on the latest period before $T_\mathrm{in} $, where sterile neutrinos are already in equilibrium but Majorana masses do not violate lepton number fast enough to yield null neutrino chemical potentials.
Equations (\ref{qei})-(\ref{dBdt}) can be written in the form 
\begin{equation}\label{dQdT}
\frac{d Q_\chi}{d \,T^{-1}} = \alpha_\chi (B-L) \cdot 10^{12} \,\mathrm{GeV}
		\;,
\end{equation}
where the coefficients $\alpha_\chi$ are constant in a radiation dominated Universe but depend on the all set of chemical potential constraints.
Assuming that $T_\mathrm{in} $ lies above $10^{13} \,\mathrm{GeV}$ one obtains for $Q_b$ and the baryon number,
$\alpha_b \approx 0.48$ and $\alpha_B \approx 0.24$,
respectively, while $\alpha_{\tau} \approx 0$ because 
$\eta_{l_i} - \eta_{\phi} - \eta_{e_i} \approx 0$.
The integration in time gives asymmetries per unity of entropy equal to
\begin{equation}\label{qchi}
Q_\chi \approx 2 \cdot 10^{-8} \, \alpha_\chi 
\sum_{ab} 
\frac{\mathrm{Im} [(h^\dagger h)_{ba}]^2}{[(h^\dagger h)_{aa}]^2}
\frac{M_a M_b}{T_a T} g\left( \frac{T_a}{T} \right)
		\; ,
\end{equation}
where $g(x)= 2 + (x + 4 + 6/x ) e^{-x} -6/x$.

The $Q$ asymmetries grow with 
$T^{-1}$ ($g \rightarrow 2 $) 
down to the temperature $T_\mathrm{in}  < T_a$
where lepton number starts to be rapidly violated.
Then, a new constraint is enforced,
$\eta_{l_i} + \eta_\phi =0$, 
that would cause $B-L$ and all the other asymmetries to vanish if there were no decoupled $Q$ asymmetries. 
$B-L$ falls down to the level of the dominant right-handed fermion asymmetries, or total baryon number, and is constrained to be a linear combination of them.
They remain marginally damped as long as the respective Yukawa processes or weak sphaleron processes are slow, as implied by equations (\ref{qei})-(\ref{dBdt}).
As the temperature drops down a new chemical potential constraint is enforced whenever any of these interactions enters in equilibrium and $B-L$ falls down further becoming a linear combination of the remanescent set of decoupled asymmetries.
The final $B-L$ asymmetry is determined by the $Q$ asymmetries that are still approximately conserved when the $B-L$ violating, $\Delta L =2$, processes become negligible. 
From then on $B-L$ is exactly conserved. 
Assuming that this happens above $2 \cdot 10^{12} \,\mathrm{GeV}$, 
the $b_R$ Yukawa and weak sphalerons are still out of equilibrium and the final $B-L$ asymmetry is dominated by the $Q_b$ and $B$ contributions:
\begin{equation}\label{bmlstar}
B- L = \frac{1}{44} \,( 23\, B + 72\, Q_b - 2 (L_{e_R} + L_{\mu_R} + L_{\tau_R}))^{\star}
		\;.
\end{equation}
The stars mean that $Q^{\star}_b$, $B^{\star}$ and $L_{e_i}^{\star}$ are calculated from equation (\ref{qchi}) at the temperature $T_\mathrm{in}  \sim 10\, M_3$ and remain practically constant after that.
The final $B-L$ asymmetry is
\begin{equation}\label{BmLnow}
B-L \approx 2 \cdot 10^{-8} \, 
\sum_{ab} 
\frac{\mathrm{Im} [(h^\dagger h)_{ba}]^2}{[(h^\dagger h)_{aa}]^2}
\frac{M_a M_b}{T_a T_\mathrm{in} } g\left( \frac{T_a}{T_\mathrm{in} } \right)
		\; . 
\end{equation}

As the temperature goes down and all right-handed leptons and quarks gradually enter in equilibrium $B-L$ is redistributed among particles to eventually yield
the present baryon number asymmetry
$B_0 \approx \frac{1}{3} (B-L)$~\cite{Harvey90,Khlebnikov88,Khlebnikov96}.
One obtains $B-L$ and baryon asymmetries at
exactly the observed order of magnitude, 
$B-L \gtrsim 10^{-10}$ per unit of entropy,
taking precisely the prompt decay ratio values indicated by the atmospheric neutrino evidence, $K_a =\Gamma_a/H \sim 10^2$,
without assuming any particular hierarchy between Yukawa couplings.

In the solution described above $B-L$ is mainly originated from the baryon number and  $b_R$ asymmetry $Q_b = B_{b_R} - B_{u_R}$. 
It assumes that $\Delta L = 2$ reactions go out of equilibrium before the $b_R$ and weak sphaleron processes enter in equilibrium~\cite{Bento03} at
$T \approx 2 \cdot 10^{12} \,\mathrm{GeV}$.
As equation (\ref{tout}) indicates this requires that the light neutrino mass scale $\bar{m}^2$ is not larger than $0.2 \,\mathrm{eV}^2$, which is still one order of magnitude higher than the atmospheric neutrino mass gap.
If $\bar{m}^2$ goes beyond $0.2 \,\mathrm{eV}^2$, the charm $c_R$ replaces $b_R$ in the role of protecting $B-L$ but $Q_c^\star \approx 0.03 \, Q_b^\star$ is too small unless it is compensated by a strong hierarchy in the Yukawa coupling structure.

\section{Conclusions}

To conclude, leptogenesis can operate in a regime where all sterile neutrinos,
including the lightest one,
decay promptly when they vanish from the Universe.
The light neutrino mass scale $\bar{m}^2 = \sum m_\nu^2$
determines the natural order of magnitude of the final baryon asymmetry.
First, $\bar{m}^2$ values between the atmospheric neutrino mass gap and 
$0.2 \,\mathrm{eV}^2$
imply that the heavy neutrinos decay promptly with ratios 
$K_a = \Gamma_a /H \sim 10^{2}$,
and reach thermal equilibrium at temperatures 10 times or more above their masses.
During thermal production, $B-L$ is generated in both lepton and sterile neutrino sectors reaching the $10^{-7}$ level.
Later, when the Majorana masses rapidly violate lepton number, $B-L$ is effectively protected by the chemically decoupled $b_R$ quark and baryon number sector if, again, 
$\bar{m}^2 \lesssim 0.2 \,\mathrm{eV}^2$, provided that the Majorana masses lie above the $b_R$ and weak sphaleron
equilibrium temperature around $2 \cdot 10^{12} \,\mathrm{GeV}$.
Then, $B-L$ falls down 3 orders of magnitude, which meets exactly the observed baryon number asymmetry $B$ without any further conditions.

The dependence of $B-L$ on the parameters of the theory is totally different from the delayed decay scenario.
In the usual $M_1 \ll M_2 ,M_3$ hierarchical case,
the lepton asymmetry generated in $N_1$ decays vanishes in the limit of degenerate light neutrinos and~\cite{davidson02,Buchmuller02} is essentially proportional to the atmospheric neutrino mass gap and the imaginary part 
$\mathrm{Im}[O_{31}^2 ]$, where $O_{31}$ is an element of the orthogonal complex matrix
$O = i v\, m_\nu^{-1/2} U \, h\, M^{-1/2}$ 
that parametrizes the couplings $h_{ia}$, and 
$m_\nu = U\, m\, U^T$ is the diagonalized neutrino mass matrix.
On the contrary, in the thermal production regime $B-L$ does not vanish in the degenerate limit and goes with 
$\mathrm{Im}[(O^\dagger O)_{23}^2 ]$.
Equations (\ref{qchi}) and (\ref{bmlstar}) yield $b_R$ and $B-L$ asymmetries of the required $10^{-10}$ order for a wide range of parameters. They only depend on light and heavy neutrino mass ratios and  matrix $O$. 

\medskip
This work was partially supported by the FCT grant CERN/FIS/43666/2001.

%


\begin{thebibliography}{28.}
\addcontentsline{toc}{section}{References}

\bibitem{Buchmuller00}
W.~Buchm\"{u}ller, M.~Pl\"{u}macher: 
Int.\ J.\ Mod.\ Phys.\  A \textbf{15}, 5047 (2000);
Phys.\ Rep.\  \textbf{320}, 329 (1999)

\bibitem{Fukugita86} 
M.~Fukugita, T.~Yanagida:
 Phys.\ Lett.\ B\  \textbf{174}, 45 (1986) 

\bibitem{Kuzmin85} 
V.~Kuzmin, V.~Rubakov, M.~Shaposhnikov: 
Phys.\ Lett.\  \textbf{155B}, 36 (1985) 

\bibitem{Gellmann79}  M.~Gell-Mann, P.~Ramond, R.~Slansky. 
In: \emph{Supergravity, Proceedings of the Workshop, Stony Brook, N.\ Y., 1979}, 
ed.\ by P.~van Nieuwenhuizen, D.~Freedman 
(North-Holland, Amsterdam, 1979)
T.~Yanagida. 
In: \emph{Proceedings of the Workshop on Unified Theories and Baryon Number in the Universe, Tsukuda, Japan, 1979}, 
ed.\ by A.~Sawada, A.~Sugamoto 
(KEK Report No. 79-18, Tsukuda, 1979)
T.~Yanagida: 
Prog. Th. Phys. B \textbf{135}, 66 (1979) 

\bibitem{Sakharov67} A.\ D. Sakharov:
JETP Lett.\ \textbf{5}, 24 (1967)

\bibitem{Fischler91} W.~Fischler, G.~F. Giudice, R.~G. Leigh, S.~Paban:
Phys.\ Lett.\ B\  \textbf{258}, 45 (1991)
W.~Buchm\"{u}ller, Y.~Yanagida:
\textit{ibid.} \textbf{302}, 240 (1993)

\bibitem{skatm00}  S.~Fukuda \textit{et~al.}, Super-Kamiokande
Collaboration:
Phys.\ Rev.\ Lett.\  {\textbf{85}}, 3999 (2000)

\bibitem{davidson02} S.~Davidson, A.~Ibarra:
Phys.\ Lett.\ B\  \textbf{535}, 25 (2002)

\bibitem{Buchmuller02}
J.~Ellis, M.~Raidal:
Nucl.\ Phys.\  \textbf{B643}, 229 (2002)
W.~Buchm\"{u}ller, P.~Di Bari, M.~Pl\"{u}macher:
\textit{ibid.}  \textbf{B643}, 295 (2002); hep-ph/0209301

\bibitem{Kolb90}
E.~W. Kolb, M.~S. Turner:
\textit{The Early Universe}, (Addison-Wesley, Reading, MA, 1990)

\bibitem{Bento03}
L.~Bento:
hep-ph/0304263

\bibitem{Fukugita90}
M.~Fukugita, T.~Yanagida:
Phys.\ Rev. D\  \textbf{42}, 1285 (1990)

\bibitem{Harvey90}
J.~A. Harvey, M.~S. Turner:
Phys.\ Rev. D\  {\textbf{42}}, 3344 (1990)

\bibitem{Cline93}
J.~M. Cline, K.~Kainulainen, K.~A. Olive:
Phys.\ Rev.\ Lett.\  \textbf{71}, 2372 (1993); Phys.\ Rev. D\  \textbf{49}, 6394 (1994)
B.~A. Campbell, S.~Davidson, J.~Ellis, K.~A. Olive:
Phys.\ Lett.\ B\  \textbf{297}, 118 (1992)

\bibitem{Weinberg79}
S.~Weinberg:
Phys.\ Rev.\ Lett.\  \textbf{42}, 850 (1979)

\bibitem{Kolb80}
E.~W. Kolb, S.~Wolfram:
Nucl.\ Phys.\  \textbf{B172}, 224 (1980); 
\textbf{B195}, 542 (E) (1982)
 
\bibitem{bento02} L.~Bento: in preparation. 
 
\bibitem{Davidson94}
S.~Davidson, K.~Kainulainen, K.~A. Olive: 
Phys.\ Lett.\ B\  \textbf{335}, 339 (1994)

\bibitem{McLerran91}
L.~McLerran, E.~Mottola, M.~E. Shaposhnikov:
Phys.\ Rev. D\  \textbf{43}, 2027 (1991)
R.~N. Mohapatra, X.~Zhang:
Phys.\ Rev. D\  \textbf{45}, 2699 (1992)

\bibitem{Khlebnikov88}  S.~Yu. Khlebnikov, M.~E. Shaposhnikov:
Nucl.\ Phys.\  \textbf{B308}, 885 (1988)

\bibitem{Bochkarev87}
A.~I. Bochkarev, M.~E. Shaposhnikov:
Mod.\ Phys.\ Lett.\ A\   \textbf{2}, 417~(1987)

\bibitem{Dine90}
M.~Dine, O.~Lechtenfeld, B.~Sakita:
Nucl.\ Phys.\  \textbf{B342}, 381~(1990)

\bibitem{Khlebnikov96}
S.~Yu. Khlebnikov, M.~E. Shaposhnikov:
Phys.\ Lett.\ B\  \textbf{387}, 817~(1996)
M.~Laine, M.~Shaposhnikov:
Phys.\ Rev. D\  \textbf{61}, 117302~(2000)

\end{thebibliography}
\end{document}